\documentclass[aps, 10pt, prL,
notitlepage, twocolumn, superscriptaddress,
nofootinbib]{revtex4-1}

\usepackage[linktocpage,breaklinks]{hyperref}
\usepackage[usenames,dvipsnames]{xcolor}

\usepackage[T1]{fontenc}
\usepackage{amsmath,times}
\usepackage{tensor}
\usepackage[utf8]{inputenc}
\usepackage{mathrsfs}
\usepackage{bm}

\usepackage{graphicx}
\usepackage{epsfig}
\usepackage{epstopdf}

\usepackage{natbib}
\usepackage{hyperref}
\usepackage{cleveref}
\definecolor{coolblack}{rgb}{0.0, 0.18, 0.39}
\definecolor{darkred}{rgb}{0.5,0,0}
\definecolor{darkgreen}{rgb}{0,0.5,0}
\definecolor{darkblue}{rgb}{0,0,0.5}
\definecolor{lapislazuli}{rgb}{0.15, 0.38, 0.61}
\definecolor{venetianred}{rgb}{0.78, 0.03, 0.08}
\definecolor{bleudefrance}{rgb}{0.19, 0.55, 0.91}
\definecolor{dogwoodrose}{rgb}{0.84, 0.09, 0.41}
\definecolor{dogwoodrose}{rgb}{0.84, 0.09, 0.41}
\hypersetup{colorlinks=true, citecolor=darkblue, linkcolor=darkblue, 
urlcolor = darkblue}

\newcommand{\sigabs}{\sigma_{\rm abs}}

\usepackage{amsmath,amssymb}
\usepackage{tensor}

\newcommand{\bea}{\begin{eqnarray}}
\newcommand{\eea}{\end{eqnarray}}
\def\l{\left}
\def\r{\right}
\def\be{\begin{equation}}
\def\ee{\end{equation}}

\begin{document}

\title{\large Spectral lines of extreme compact objects}

\author{Caio F. B. Macedo}%\email{caiomacedo@ufpa.br}
%\affiliation{Faculdade de F\'{\i}sica, Universidade 
%Federal do Par\'a, 66075-110, Bel\'em, Par\'a, Brazil.}
\affiliation{Campus Universit\'ario Salin\'opolis, Universidade 
Federal do Par\'a, 68721-000, Salinópolis, Par\'a, Brazil.}

\author{Tom Stratton}%\email{tstratton1@sheffield.ac.uk}
\affiliation{Consortium for Fundamental Physics, School of Mathematics and Statistics,
University of Sheffield, Hicks Building, Hounsfield Road, Sheffield S3 7RH, United Kingdom}

\author{Sam Dolan}%\email{s.dolan@sheffield.ac.uk}
\affiliation{Consortium for Fundamental Physics, School of Mathematics and Statistics,
University of Sheffield, Hicks Building, Hounsfield Road, Sheffield S3 7RH, United Kingdom}

\author{Lu\'is C. B. Crispino}%\email{crispino@ufpa.br}
\affiliation{Faculdade de F\'{\i}sica, Universidade 
Federal do Par\'a, 66075-110, Bel\'em, Par\'a, Brazil.}

\begin{abstract}

We study the absorption of scalar fields by extreme/exotic compact objects (ECOs) -- horizonless alternatives to black holes -- via a simple model in which dissipative mechanisms are encapsulated in a single parameter. Trapped modes, localized between the ECO core and the potential barrier at the photonsphere, generate Breit-Wigner-type spectral lines in the absorption cross section. Absorption is enhanced whenever the wave frequency resonates with a trapped mode, leading to a spectral profile which differs qualitatively from that of a black hole. We introduce a model based on Nariai spacetime, in which properties of the spectral lines are calculated in closed form. We present numerically calculated absorption cross sections and transmission factors for example scenarios, and show how the Nariai model captures the essential features. We argue that, in principle, ECOs can be distinguished from black holes through their absorption spectra.

\end{abstract}


\date{\today}

\maketitle

%\tableofcontents

\section{Introduction}

 The recent detections of gravitational waves (GWs) have reinforced the position of general relativity (GR) as the canonical theory of gravity~\cite{Abbott:2016blz,Abbott:2016nmj,TheLIGOScientific:2016wfe,TheLIGOScientific:2016htt}. 
In the GW150914 event, the loudest thus far, no significant evidence for 
violations of GR has been found \cite{TheLIGOScientific:2016src,Yunes:2016jcc}, and the 
dynamics appears fully consistent with the coalescence of two black holes (BHs). In 2017, alternative theories of gravity were strongly constrained by the near-coincident arrival of GWs and gamma rays from a binary neutron star inspiral \cite{TheLIGOScientific:2017qsa}.

GW signals probe the spacetime up to the photonsphere, rather than the event horizon itself, it has been argued \cite{Cardoso:2016rao}. The possibility lingers that the progenitors of e.g.~GW150914 are extreme/exotic compact objects (ECOs) which mimic properties of BHs. The next decade will see a concerted effort to address the question of whether event horizons truly form in nature; and whether horizons are ``clean'', i.e.,~free from noncanonical features such as firewalls \cite{Almheiri:2012rt}. This effort necessitates a clear understanding of the generic properties of horizonless alternatives to BHs.  

The standard picture for the evolution of BH binaries is divided into three main stages: {\it (i)} inspiral, {\it (ii)} merger, and {\it (iii)} ringdown. %The inspiral phase, in which the binaries have large radii, is essentially described by post-Newtonian perturbative corrections~\cite{Blanchet2014}. The merger phase, in which the binaries evolve to very small radii and high velocities, is the domain of full numerical relativity. 
The ringdown signal can be modeled through a combination of the final object modes, known as quasinormal modes~\cite{Berti:2005ys}. The ringdown phase for signals with large a signal-to-noise ratio 
can shed light on the nature of the remnant compact objects, and on gravity itself~\cite{Berti:2015itd}.

There are several proposals for alternatives to BHs, that nevertheless produce ringdown signals that closely mimic those of BHs in GR at early times. To assess these alternatives, it is necessary to analyze the subtle differences between the signatures of BH mimickers and true BHs~\cite{Cardoso:2016rao}. 
Generically, compact horizonless objects ($R<3M$) possess long-lived modes, which are related to trapped $w$ modes \cite{Kokkotas:1999bd,Kokkotas:1994an,Cardoso:2014sna}. These modes are associated with a minimum in the effective potential, that (in the eikonal limit) corresponds to a stable null geodesic present within the stellar configuration~\cite{Cardoso:2014sna}. In GW binaries, long-lived modes are expected to leave imprints in the phenomenology, most notably in resonant configurations~\cite{Macedo:2013jja,Macedo:2013qea,Kojima:1987tk}. 

ECOs may arise via near-horizon modifications of gravitational collapse~\cite{Barcelo:2015noa,Barcelo:2017lnx} or as exotic solutions such as gravastars~\cite{Mazur:2001fv} or boson 
stars~\cite{Schunck:2003kk,Macedo:2013jja}.
ECO candidates are classified as either ultracompact objects (UCOs) or clean 
photonsphere objects (ClePhOs)~\cite{Cardoso:2017njb}. UCOs are compact enough that the 
spacetime presents a photonsphere ($R < 3M$). ClePhOs possess not only a photonsphere but also a spectrum of  
modes trapped within it that may provide a clean signal. There has been much recent interest in searches for evidence of echoes from ECOs and ultracompact objects in gravitational-wave data \cite{Abedi:2016hgu, Akhmedov:2016uha, Westerweck:2017hus, Mark:2017dnq, Conklin:2017lwb,Barausse:2018vdb}. 

\begin{figure}%
\includegraphics[width=\columnwidth]{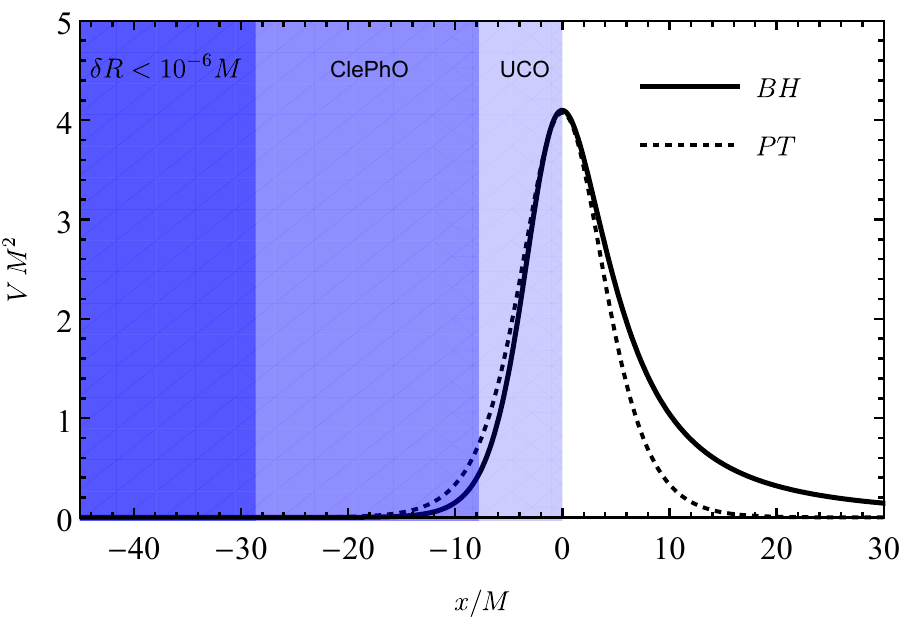}%
\caption{Effective potential for wave propagation in compact stars. 
The Schwarzschild BH [solid] and P\"oschl-Teller (PT) [dashed] potentials are shown as a function of tortoise coordinate $x$ (\ref{eq:tortoise}). Shaded regions indicate the surface radii for classes of compact stars.
%See also Ref.~\cite{Cardoso:2016oxy}.
}%
\label{EP}%
\end{figure}

As Fig.~\ref{EP} illustrates, the key characteristic of a ClePhO is an effective ``cavity'' in the high-redshift region between the object's surface and the maximum of the potential barrier defining the photonshere. For static objects of mass $M$, the cavity width is characterised by the ``tortoise coordinate'' of the surface, $x_0 = x(R)$, with 
\be
x(r) \equiv r + r_h \ln(r/r_h - 1) + \kappa  , \label{eq:tortoise}
\ee
choosing the constant $\kappa = - (3-2\ln2) M$ such that the peak of the potential is at $x = 0$. We adopt units $G=c=1$, such that the event horizon lies at $r = r_h =2M$ and the surface lies at $R = r_h + \delta R$. 
The cavity is associated with long-lived modes that correspond to poles of the scattering matrix $\bf{S}$~\cite{Berti:2009kk,Berti:2009wx,Pani:2013pma}. 

Heuristically, the width of the cavity determines both the spacing of the trapped modes in the frequency domain $\Delta \omega$ [see Eq.~(\ref{eq:wapprox})] and the delay between echoes $\Delta t = 2 \pi / \Delta \omega$ in the time domain. A surface/firewall at a proper distance \cite{Abedi:2016hgu} from the horizon of the Plank length $l_P$ yields $\delta R = l_P^2 / (4r_h)$ and $|x_0|/M \approx 4 \ln (l_P / r_h) \sim 366 + 4 \ln( \frac{M}{60M_{\odot}})$; thus $\Delta \omega \sim 4.5 \, \text{Hz} \times \frac{60M_\odot}{M}$ and $\Delta t \sim 0.22 \,\text{s} \times \frac{M}{60M_\odot} $.

In this work we highlight a key observational signature of a ClePhO that would unambiguously distinguish it from a true BH. The absorption cross section of a ClePhO is characterized by spectral lines that are reminiscent of atomic/molecular absorption lines. These lines arise directly from the trapped-mode spectrum $\{\omega_{l n}\}$. An observation of absorption lines would reveal not just the width of the ClePhO cavity, but also the degree of dissipation at the ClePhO surface (or firewall). We show below that the modal transmission amplitudes $\Gamma_{\omega l}$ bear the imprint of Breit-Wigner-type resonances \cite{1991RSPSA.434..449C}, familiar from e.g.~nuclear scattering theory \cite{breit1936capture,feshbach1947scattering}, viz.,
\be
\Gamma_{\omega l} \approx \sum_{n=0}^\infty \frac{\mathcal{A}_{l n}}{(\omega - \text{Re} \, \omega_{l n})^2 + (\text{Im} \, \omega_{l n})^2} .
\label{eq:BWig}
\ee
Below we obtain closed-form approximations for the mode spectrum $\omega_{ln}$ and amplitudes $\mathcal{A}_{l n}$; and we compute the absorption cross section numerically for illustrative scenarios. \\

\section{Model and assumptions}

Spherically symmetric spacetimes can be generically described by the line element
\be
ds^2=-A(r) dt^2+B(r)^{-1}dr^2+r^2d\Omega^2.
\ee
%The functions $A(r)$ and $B(r)$ are determined by solving the field equations for a given theory within the star. 
In the standard picture, GR is assumed to be valid outside the compact object, with the solution given by the Schwarzschild spacetime, i.e., $A(r)=B(r)=1
-2M/r$, where $M$ is the total mass of the ECO. The inner structure of the spacetime depends on the model; 
 this can be translated into a boundary 
condition on the field at the object's surface.

The field may be written in separable form as
\be
\Phi = \sum_{l=0}^\infty a_{\omega l} \frac{u_{\omega l}(r)}{r}Y_{l0}(\theta,\varphi)e^{-i\omega t},
\label{eq:wave}
\ee
where $a_{\omega l}$ are coefficients. 
We choose boundary conditions such that, far from the ECO, $\Phi$ is the sum of a (distorted) planar wave and an outgoing scattered component \cite{Crispino:2009ki}. 

The scalar field is governed by the Klein-Gordon equation $\Box_g \Phi = 0$. Inserting Eq.~(\ref{eq:wave}) leads to a radial equation
\be
\l[ \frac{d^2}{dx^2}+\omega^2 - V(r) \r] u_{\omega l} = 0, \; V \equiv A\l[\frac{l(l+1)}{r^2}+\frac{(AB)'}{2Ar}\r] .\label{eq:kgradial}
\ee
Figure \ref{EP} shows the effective potential for the Schwarzschild spacetime.

For a compact object, each mode admits a regular power series expansion near its center, $u_{\omega l} \sim r^{l+1} \sum b_k r^{2k}$. Typically, one extends the solution---via numerical integration or otherwise---to the surface at $r=R$, and then extracts $[u_{\omega l}(R),u'_{\omega l}(R)]$ to place a boundary condition on the Schwarzschild exterior. Here, instead, we shall start at the surface with a heuristic boundary condition, allowing us to draw a veil over the (model-dependent and unknown) interior structure. For a ClePhO, $R = r_h + \delta R$, where $\delta R \ll r_h$, and thus $V(R) \approx 0$. Thus, we may write
\be
u_{\omega l}(r \approx R) \approx e^{-i \omega x} + \left( \mathcal{K}  e^{-2 i\omega x_0} \right) e^{+i \omega x}, \label{eq:cond}
\ee
where $\mathcal{K}$ is a parameter characterising the reflectivity of the body \cite{Maggio:2017ivp}. 
An advantage of this parametrization is that we may impose Dirichlet-type (${\cal K }=-1$), Neumann-type (${\cal K }=1$), or BH-type (${\cal K}=0$) boundary conditions through a single parameter ${\cal K}$. 

If the field interacts directly with the ECO itself, by inducing bulk motion or through some nontrivial coupling, this will likely lead to dissipative frictionlike effects. Thus, we consider ECOs with $|{\cal K}|<1$, with $|{\cal K}|\approx 1$ for weakly interacting media. Our aim is to build a heuristic understanding of the role of dissipation, without a specific model of the underlying physics. %(In general, ${\cal K }$ and $x_0$ could vary with $\omega$, but we do not consider that case here).

\vspace{0.5cm}

\section{Absorption by an ECO}

The absorption cross section for a spherical-symmetric object is 
\be
\sigma_{\rm abs}(\omega) = \frac{\pi}{\omega^2}\sum_{l}(2l+1) \Gamma_{\omega l} ,
\label{eq:sigabs}
\ee
where $\Gamma_{\omega l}$ are modal transmission factors {(proportional to the square modulus of the ``transfer functions'' of Ref.~\cite{Mark:2017dnq,Conklin:2017lwb})} 
defined by 
\be
\Gamma_{\omega l} \equiv 1 - \left| \frac{A^+_s}{A^-_s} \right|^2 = \frac{1 - |\mathcal{K}|^2}{|A^-_s|^2} . \label{eq:Gamma}
\ee
Here $A^\pm_s$ are modal constants obtained from solutions of the radial equation (\ref{eq:kgradial}) obeying % the ECO boundary conditions,
\be
u^{s}_{\omega l} = \begin{cases}
e^{-i \omega x} + \mathcal{K} e^{-2i\omega x_0} e^{+i \omega x}  , & x \rightarrow x_0 , \\
A_s^- e^{-i\omega x} + A_s^+ e^{+i\omega x}, & x \rightarrow \infty ,
\end{cases}
\label{eq:us-bc}
\ee
where $x_0$ defines the surface and $x$ is defined in Eq.~(\ref{eq:tortoise}).

The mode $u^s_{\omega l}$ is a linear combination of the standard ``in'' and ``up'' solutions of the BH scattering problem, 
viz.,
\begin{subequations}
\label{eq:uinuup}
\begin{align}
u^{in}_{\omega l} &= \begin{cases}
e^{-i \omega x} , & x \rightarrow -\infty , \\
A_\infty^- e^{-i\omega x} + A_\infty^+ e^{+i\omega x}, & x \rightarrow +\infty ,
\end{cases} \\
u^{up}_{\omega l} &= \begin{cases}
A_h^- e^{-i \omega x} + A_h^+ e^{+i \omega x} , & x \rightarrow -\infty , \\
\qquad \qquad \qquad \;\, e^{+i\omega x}, & x \rightarrow +\infty .
\end{cases}
\end{align}
\end{subequations}
From the Wronskian relations between $u^{in}_{\omega l}$, $u^{up}_{\omega l}$ and $u^s_{\omega l}$, it follows that $A_h^+ = A_{\infty}^-$, $A_s^+ A_\infty^- - A_s^- A_\infty^+ = \mathcal{K} e^{-2 i \omega x_0}$ and 
\be
A_s^- = A_h^+ - A_h^- \mathcal{K} e^{-2 i \omega x_0} . \label{eq:Asm}
\ee 
By inserting Eq.~(\ref{eq:Asm}) into Eq.~(\ref{eq:Gamma}), one may compute ClePhO transmission factors directly from the standard BH up-mode coefficients. 

The transmission factors $\Gamma_{\omega l}$ are singular where $A_s^-$ is zero, i.e., where
\be
A_h^+ / A_h^- = \mathcal{K} e^{-2 i \omega x_0} . \label{eq:qnm-condition}
\ee
This condition defines a spectrum of (complex) modes $\{ \omega_{ln} \}$ for a compact body.

Close to a trapped-mode frequency $\omega_{ln}$, one has $A_s^- \approx (\omega - \omega_{ln}) \left. \partial_\omega  A_s^- \right|_{\omega=\omega_{ln}}$. 
By using Eq.~(\ref{eq:Gamma}), it follows, for real frequencies $\omega$, that $\Gamma_{\omega l}$ takes the Breit-Wigner form (\ref{eq:BWig}), with  $\mathcal{A}_{ln} = (1 - \mathcal{K}^2) \left| \partial_\omega A_s^- \right|^{-2}$. 
For insight into the spectrum and the amplitude $\mathcal{A}_{ln}$, we now turn to an approximate mode and then numerical methods. \\

 %For comparison, the black hole quasinormal modes are defined by $A_h^+ / A_h^- = 0$. \\

\section{The comparison problem: Nariai spacetime} 

We now consider the Nariai spacetime ($dS_2 \times S_2$) in which the transmission/reflection problem can be solved in closed form, with line element %This spacetime, a limit of  Schwarzschild-de Sitter, has line element
\be
d\hat{s}^2 = -F(y) d \hat{t}^2 + F^{-1}(y) d y^2 + d\Omega_2^2,
\ee
where $F(y) = 1 - y^2$ and $y \in (-1,+1)$. The Klein-Gordon equation $(\Box + \frac{1}{8}R) \Phi = 0$ generates the radial equation
\be
\left\{ \frac{d^2}{d\hat{x}^2} + \hat{\omega}^2 - \frac{L^2 + 1/4}{\cosh^2 \hat{x}} \right\} \hat{u}_{l\omega} = 0,
\ee
where $\hat{x} = \tanh^{-1} y$ and $L \equiv l+1/2$. The potential barrier, which is of P\"oschl-Teller type \cite{Poschl:1933zz}, is similar in structure to the Schwarzschild barrier (see Fig.~\ref{EP}), with a closest match for $\hat{x} =  x / \nu$, $\hat{\omega} = \nu \omega$ where $\nu \equiv \sqrt{27} M$. 
Standard solutions $\hat{u}^{in}_{l\omega}$ and $\hat{u}^{up}_{l\omega}$ are defined by analogy to Eq.~(\ref{eq:uinuup}). These are known in closed form in terms of Legendre functions \cite{Casals:2009zh}. The  coefficients are
\begin{subequations}
\label{eq:Anar}
\begin{align}
\hat{A}_h^+ &= \frac{\Gamma(-i\hat{\omega})\Gamma(1 - i\hat{\omega})}{\Gamma(\frac{1}{2} + i L- i \hat{\omega})\Gamma(\frac{1}{2} - iL - i\hat{\omega})}, \\
\hat{A}_h^- &= \frac{\Gamma(i\hat{\omega})\Gamma(1-i\hat{\omega})}{\Gamma(\frac{1}{2} + iL)\Gamma(\frac{1}{2} - iL)} .
\end{align}
\end{subequations}
%where $\Gamma(\cdot)$ denotes the Gamma function. 
As the Nariai potential is symmetric under $x \leftrightarrow -x$, it follows that $\hat{A}^\pm_h = \hat{A}^\mp_\infty$. 
An expression for the transmission factor $\hat{\Gamma}_{\omega l}$ is found by inserting Eqs.~(\ref{eq:Anar}) into Eqs.~(\ref{eq:Asm}) and (\ref{eq:Gamma}). 

The standard BH quasinormal modes are defined by $\hat{A}^-_\infty / \hat{A}^+_\infty = 0$, yielding a spectrum $\hat{\omega}_{l n} = \pm (l+1/2) - i (n+1/2)$, where $n \in \mathbb{Z}$. Conversely, compact star trapped modes are defined instead by Eq.~(\ref{eq:qnm-condition}), yielding the condition
\be
\frac{\Gamma(-i\hat{\omega})}{\Gamma(+i \hat{\omega})} \frac{\Gamma(\frac{1}{2} + i L) \Gamma(\frac{1}{2} - iL)}{\Gamma(\frac{1}{2} + iL - i \hat{\omega})\Gamma(\frac{1}{2} -iL -i\hat{\omega})}
=
\mathcal{K} e^{-2i \hat{\omega} \hat{x}_0}.
\ee
In the regime $\hat{\omega} \ll L$, the left-hand side is approximately $-1$, and (for $\mathcal{K} > 0$) the spectrum is approximated by
\be
\hat{\omega}_{l n} \approx \frac{\pi (n+1/2)}{|\hat{x}_0|}  + i \frac{\ln \mathcal{|K|}}{2 |\hat{x}_0|} , \label{eq:wapprox}
\ee
i.e., an evenly spaced spectrum of resonances with approximately constant Lorentzian width set by $\ln |\mathcal{K}|$.

To deduce the amplitude $\hat{\mathcal{A}}_{ln}$, we use $\left. \partial_{\hat{\omega}} A_s^{-} \right|_{\hat{\omega}_{ln}} = \hat{A}_h^+ \left(2 i \hat{x}_0 - \partial_{\hat{\omega}} \ln \alpha \right)$, where $\alpha \equiv \hat{A}_h^- / \hat{A}_h^+$. The former term dominates over the latter for $\hat{\omega} \ll L$. If $\text{Re} \, \hat{\omega}_{ln} \gg \text{Im} \, \hat{\omega}_{ln}$, we may evaluate $|\hat{A}^+_h|^2$ for real frequencies without substantial loss of accuracy, to obtain an expression for the amplitude in the Breit-Wigner formula (\ref{eq:BWig}),
\be
\hat{\mathcal{A}}_{ln} \approx \frac{1-\mathcal{K}^2}{4 x_0^2} \frac{\sinh^2(\pi \hat{\omega})}{\cosh(\pi(L-\hat{\omega})) \cosh(\pi(L+\hat{\omega}))} .  \label{eq:Ampl-approx}
\ee
For $L \gg \hat{\omega}$, the spectral lines are exponentially suppressed, as from Eq.~(\ref{eq:Ampl-approx}) the amplitude scales with $\exp(2 \pi \hat{\omega})$ in this regime. The spectral lines become significant for $\hat{\omega} \sim L$.

\section{Numerical method} 

We computed the absorption cross section $\sigabs$, given by Eq.~(\ref{eq:sigabs}), and the mode spectrum using numerical techniques. The task, in outline, was to compute $\Gamma_{\omega l}$ via Eq.~(\ref{eq:Gamma}) by first solving the radial equation (\ref{eq:kgradial}) to obtain the ingoing and outgoing coefficients $A^\pm_s$ in Eq.~(\ref{eq:us-bc}). 

%We applied a standard numerical-integration method to solve Eq.~(\ref{eq:kgradial}) and extract the coefficients $A^\pm_s$. 
Far from the object, the potential $V$ approaches zero (cf. Fig.~\ref{EP}), and the mode can be written as
$
u_{\omega l} \approx A^-_s \mathcal{R}_{\omega l}+ A^+_s \mathcal{R}_{(-\omega) l},\label{eq:numericalexp}
$
where 
$
\mathcal{R}_{\omega l}=e^{-i\omega r_*}\sum_{j=0}^{N} B_{(j)} r^{-j} .
$
The coefficients $B_{(j)}$ were obtained iteratively, by expanding Eq.~\eqref{eq:kgradial} in powers of $r^{-1}$ in the asymptotic region. Typically, we truncated at order $N=15$. %, guaranteeing numerical convergence of the solution. 
To obtain the numerical coefficients $A^\pm_s$, we integrated the differential equation (\ref{eq:kgradial}) from the surface of the star outward into a region where $\omega^2\gg V(r)$, then matched the numerical solution onto the asymptotic form above. \\ %(We have checked the stability of the solutions by changing the asymptotic radius in the integrations.) \\

\section{Results} 
%
%Before presenting numerical results, let us first examine the low- and high-frequency limits.
At low frequencies $M\omega \ll 1$, we find via the methods of Ref.~\cite{Unruh:1976fm} that the absorption cross section is
\be \label{eq:lowfreq}
\lim_{M\omega \rightarrow 0} \sigabs  = \text{Re} \frac{1-\mathcal{K}}{1+\mathcal{K}} \, \sigma_{\rm BH},
\ee
where $\sigma_{\rm BH} = 27 \pi M^2$. Absorption at low frequencies occurs, for example, by fluid accretion onto a moving object \cite{Petrich:1988zz}. At high frequencies, $\sigabs$ fluctuates around the value
\be
\sigabs \sim \sigma_{\rm BH}(1-|\mathcal{K}|^2).
\label{eq:high}
\ee
Between these limits, there is significant structure in $\sigabs$. 
%Equation (\ref{eq:high}) tends to the Schwarzschild limit for ${\cal K}=0$.

%The approximation \eqref{eq:lowfreq} is valid for $\mathcal{K} \neq -1$, and it reduces to the Schwarzschild result when $\mathcal{K} =0$, as expected. Additionally, in the high-frequency limit the absorption cross section fluctuates around the following value
%\be
%\sigabs\approx \sigma_{\rm BH}(1-\mathcal{K}^2).
%\label{eq:high}
%\ee
%Equation (\ref{eq:high}) tends to the Schwarzschild limit for ${\cal K}=0$.

\begin{figure}[h]%
\includegraphics[width=\columnwidth]{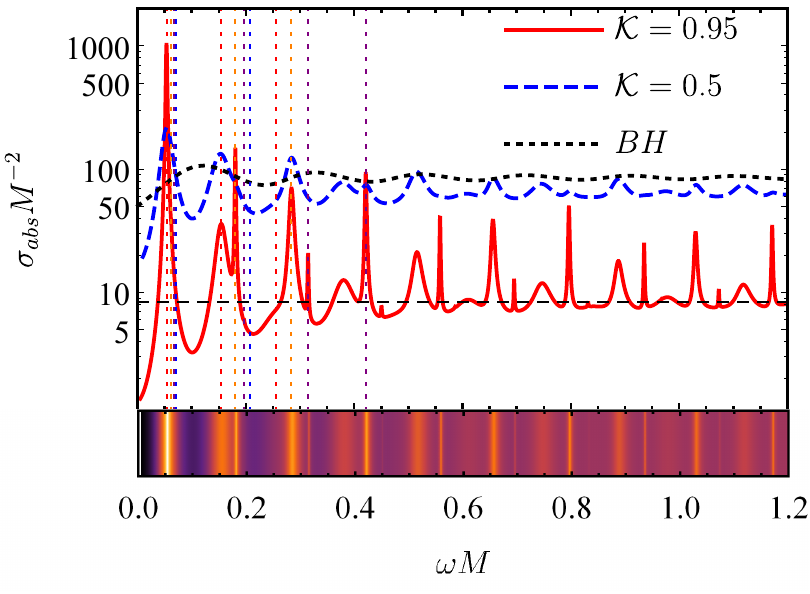}%
\caption{Absorption spectrum of an ECO with $\delta R=10^{-6}M$, with ${\cal K}=0.95$ and ${\cal K}=0.5$, and the BH case $({\cal K}=0)$. % Spectral lines in the absorption spectrum appear at the trapped-mode frequencies of the object. 
The dotted vertical lines indicate the trapped-mode frequencies of Table~\ref{tab:modes}. The absorption band below the plot is for the case ${\cal K}=0.95$.}%
\label{fig:abs}%
\end{figure}

Figure \ref{fig:abs} shows the absorption cross section of ECOs with $\delta R=10^{-6}M$ for mild (${\cal K}=0.95$) and strong (${\cal K}=0.5$) dissipative effects. In both cases, $\sigabs$ exhibits distinct peaks---i.e.~Breit-Wigner-type spectral lines---that are absent in the BH scenario (${\cal K}=0$). The spectral lines arise at frequencies set by the real part of the trapped-mode frequencies, as illustrated in Fig.~ \ref{fig:abs}. The width of the spectral lines is determined by the imaginary part of $\omega_{ln}$, and thus the dissipativity of the ECO [see Eq.~(\ref{eq:wapprox})\, with narrowing lines as ${\cal K} \rightarrow 1$. 

Figure \ref{fig:abs} shows that absorption is enhanced whenever the incoming wave excites a $w$-mode resonance significantly. Remarkably, absorption by a weakly dissipative ECO (e.g.~${\cal K}=0.95$) can exceed the absorption by a Schwarzschild BH of the same mass, if a low-frequency trapped mode is excited. The $l=0$ mode shows the strongest effect. The first few trapped-mode frequencies are enumerated in Table~\ref{tab:modes}. % For higher frequencies, the structure of the absorption spectrum remains qualitatively the same, with the occurrence of narrow peaks.

\begin{table}%
\caption{Spectrum of an ECO for $\delta R=10^{-6}M$ and $\mathcal{K}=0.95$.}
\begin{tabular}{lll | lll}
\hline\hline
$l$& $\omega_R$ & $-\omega_I$ &
$l$& $\omega_R$ & $-\omega_I$\\
\hline
0 & 0.05357 & $1.452\times 10^{-3}$ &
	1 & 0.06244 & $1.011\times 10^{-3}$\\
& 0.1544        & $1.153\times 10^{-2}$ &
	& 0.1806 & $1.061\times 10^{-3}$\\  
& 0.2563       &  $3.532\times 10^{-2}$ &
  	& 0.2841 & $4.702\times 10^{-3}$\\
&&&
	& 0.3791  & $2.264\times 10^{-2}$\\ \hline
2 & 0.0678  & $1.096\times 10^{-3}$ &
	3 & 0.07199 & $1.161\times 10^{-3}$\\
  & 0.1969 & $1.008\times 10^{-3}$ &
	& 0.2085  & $1.066\times 10^{-3}$\\
  & 0.3149 & $9.420\times 10^{-4}$ &
	& 0.3339 & $9.851\times 10^{-4}$\\
  & 0.4213 & $1.686\times 10^{-3}$ &
 	& 0.4504 & $9.219\times 10^{-4}$ \\ 
 & 0.5153 & $1.035\times 10^{-2}$ &
	& 0.5860 & $9.974\times 10^{-4}$\\
 & 0.6089 & $3.317\times 10^{-2}$\\
\hline\hline
\end{tabular}
\label{tab:modes}
\end{table}

\begin{figure}%
\includegraphics[width=\columnwidth]{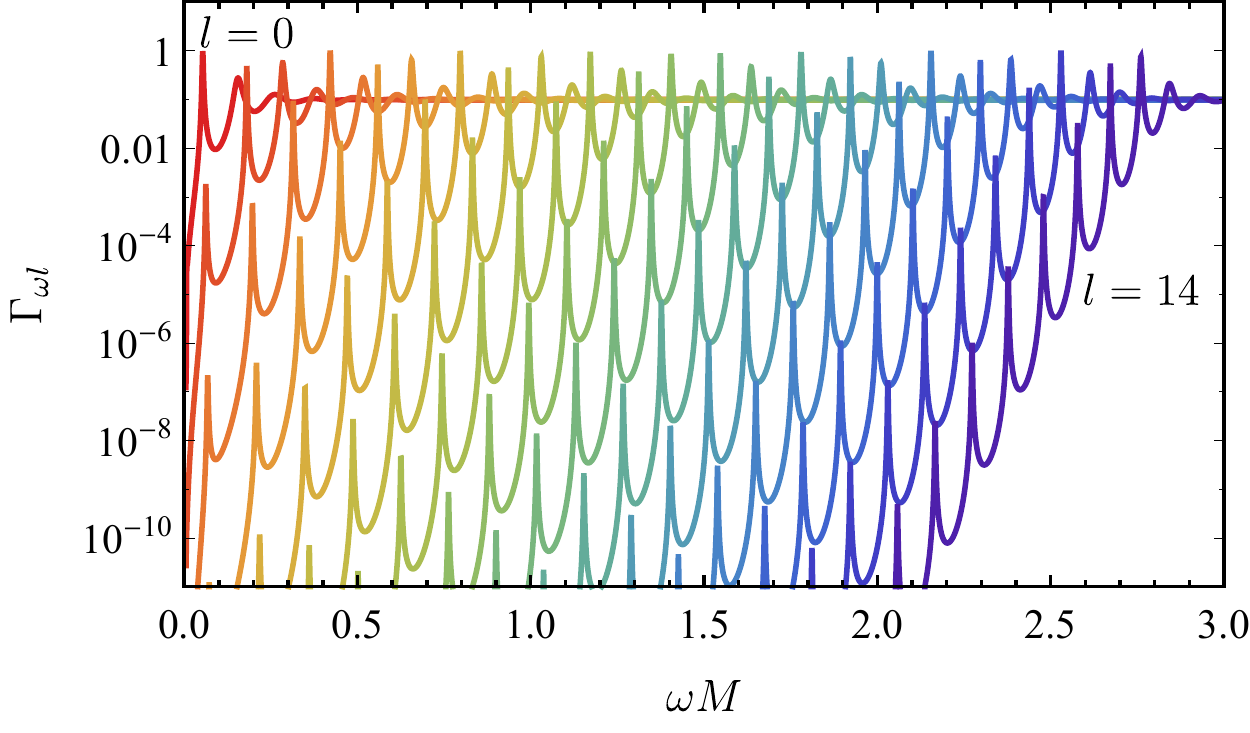}%
\caption{Partial transmission coefficients as functions of the frequency for $\delta R=10^{-6}M$ and ${\cal K}=0.95$. Peaks appear for frequencies close to the real part of the QNM frequencies. Note that the peaks persist even for higher frequencies and $l$'s.}%
\label{fig:trans}%
\end{figure}

Modal transmission factors $\Gamma_{\omega l}$ are shown in Fig.~\ref{fig:trans}, for the case $\delta R=10^{-6}M$ and ${\cal K}=0.95$, for multipoles $l = 0 \ldots 14$. For a given $l$, the transmission factor shows multiple evenly spaced Breit-Wigner spectral peaks, of approximately similar width. The amplitude of these peaks increases exponentially with $\omega$, initially, before leveling off at $1-{\cal K}^2$ for $\omega \gtrsim (l + 1/2) / \sqrt{27} M$. Once the energy $\omega^2$ exceeds the height of the potential barrier, the peaks become wider and less distinct. These qualitative features were anticipated in Eqs.~(\ref{eq:BWig}), (\ref{eq:wapprox}), and (\ref{eq:Ampl-approx}). %Additionally, narrow peaks appear for some frequencies, which are the real part of the trapped mode frequencies of the compact objects. At the peaks, the transmission coefficients can increase considerably in value and this behavior persists for higher frequencies. We shall see below that the contribution of these resonant structures has an interesting impact in the absorption cross section of these objects.

\begin{figure}%
\includegraphics[width=\columnwidth]{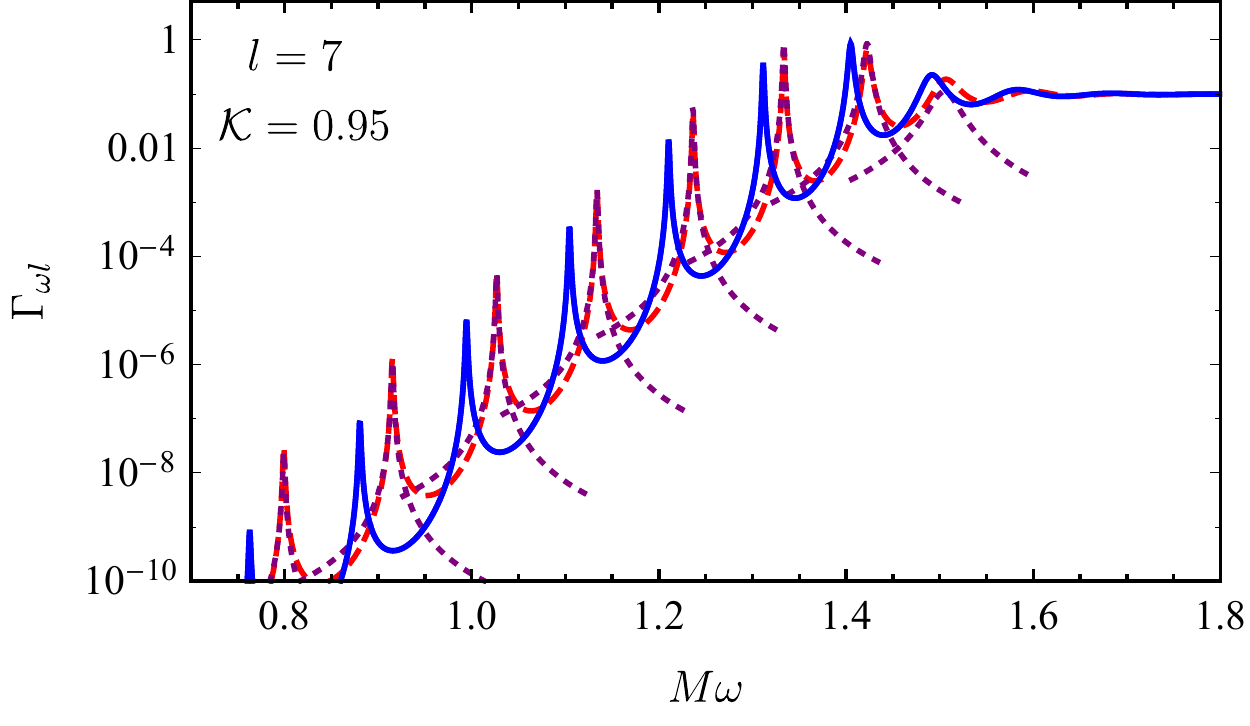}%
\caption{Transmission factors for a ClePhO (blue solid line) and for Nariai spacetime (red dashed line) for $\delta R=10^{-6}M$ and ${\cal K}=0.95$. Dotted lines show the Breit-Wigner approximation (\ref{eq:BWig}) with amplitude $\hat{\mathcal{A}}_{ln}$ (\ref{eq:Ampl-approx}).}%
\label{fig:comparison}%
\end{figure}

In Fig.~\ref{fig:comparison}, we compare the transmission factors for a ClePhO with closed-form expressions obtained for the Nariai spacetime. The plot shows that the latter serves as a robust proxy for the former and that the Breit-Wigner formula (\ref{eq:BWig}) provides a good fit to the spectral lines. Although the positions of the Nariai spectral lines do not exactly match the positions of the Schwarzschild lines, the spacing is comparable, and the amplitude approximation of (\ref{eq:Ampl-approx}) captures the essential features. %: exponential growth followed by a levelling-off at $\Gamma_{\omega l} \sim 1 - \mathcal{K}^2$. 

The transmission factors exhibit an approximate shift symmetry under $l \rightarrow l + 1$ and $\omega \rightarrow \omega + 1/\sqrt{27}M$; see Figs.~\ref{fig:trans} and \ref{fig:comparison}, and Eqs.~(\ref{eq:wapprox}) and (\ref{eq:Ampl-approx}). A consequence is that, for a given $l$, the amplitude of the dominant peak is insensitive to $l$. Hence the amplitude of the spectral lines in $\sigabs$ scales with $\omega^{-1}$ at high frequencies, allowing spectral lines to persist at frequencies substantially above $M \omega \sim 1$.

\vspace{0.5cm}

\section{Discussion and conclusion} 

We have shown that small dissipative effects in ECOs will produce Breit-Wigner-type spectral lines in absorption cross sections. We have focused on a simple model that allows parametric control over dissipation. We derived exact closed-form results for the Nariai spacetime and showed that these capture all qualitative aspects of the ClePhO case, which we studied numerically. We have argued that spectral lines are robust features in putative ECOs.  %\sd{Can we back up this statement? It seems like a big extrapolation, as neutron stars do not exhibit a photonshere.}

Spectral lines have a typical (angular-frequency) width of $-\ln|\mathcal{K}| / 2 |x_0|$ and a typical spacing of $\pi / |x_0|$ in the crudest approximation (\ref{eq:wapprox}). Individual lines are resolvable if the latter exceeds the former; that is, if $|\mathcal{K}| \gtrsim e^{-2\pi} \approx 0.002$. As Fig.~\ref{fig:abs} shows, narrow lines produced by weak dissipation ($\mathcal{K} = 0.95$) would be substantially easier to resolve than wider lines from strong dissipation ($\mathcal{K} =0.5$). 

We anticipate that the main features of scalar-field absorption will carry across to ECOs perturbed by electromagnetic ($\mathfrak{s} = 1$) and gravitational ($\mathfrak{s} = 2$) fields, with some caveats. First, the $l < \mathfrak{s}$ modes are absent in these cases. Second, these fields probe distinct frequency ranges. GWs may have a wavelength comparable to ECOs, such that $M\omega \sim 1$. On the other hand, electromagnetic waves will typically be much shorter in wavelength (e.g.~for the CMB and a solar-mass ECO, the dimensionless parameter is $M \omega \sim 10^6$). In the high-frequency limit the amplitude of the spectral lines diminishes with $1/(M\omega)$, presumably limiting their detectability. %Primordial ECOs of masses $M \ll M_{\odot}$, if extant, would produce absorption lines in EM bands. 

These results may have wider implications for known compact bodies, such as neutron stars with $R / M \sim 6$.  Although neutron stars do not exhibit a photonsphere, they do have families of fluid modes---the $f$, $p$ and $g$ modes~\cite{Kokkotas:1999bd}---with imaginary parts comparable to or smaller than those in Table~\ref{tab:modes}. In principle, spectral lines are generated by energy exchange between the impinging field and the neutron star. This gives a mechanism for enhanced accretion of weakly interacting fields, such as dark matter, and energy deposit by GWs, whenever the wave frequency mode matches a fluid mode.

Finally, we note that ECOs may also generate \emph{emission} lines in two ways. First, as $\Gamma_{\omega l}$ is a key ingredient in the Hawking radiation calculation, one might anticipate significant deviations from the near-blackbody spectrum for ECOs. Second, rotating ECOs suffer an ergoregion instability caused by superradiance \cite{Friedman:1978}, leading to the appearance of trapped modes that grow, rather than decay, with time \cite{Maggio:2017ivp}. Thus, stimulated excitation of the ergoregion instability would generate emission lines.

\section*{Acknowledgments}

The authors would like to thank Conselho Nacional de Desenvolvimento Cient\'ifico e Tecnol\'ogico (CNPq) and Coordena\c{c}\~ao de Aperfei\c{c}oamento de Pessoal de N\'ivel Superior (CAPES)- Finance Code 001, from Brazil, for partial financial support.
This research has also received funding from the European Union's Horizon 2020 research and innovation programme under the H2020-MSCA-RISE-2017 Grant No. FunFiCO-777740.
S.D.~acknowledges financial support from the Engineering and Physical Sciences Research Council (EPSRC) under Grant No.~EP/M025802/1 and from the Science and Technology Facilities Council (STFC) under Grant No.~ST/P000800/1.

\bibliography{references}

\end{document}